# A new scheme for maximizing the lifetime of heterogeneous wireless sensor networks[1]

Reem Aldaihani and Hosam AboElFotoh
Computer Science Department
College of Computing Science and Engineering
Kuwait University, Kuwait
{reem.aldaihani, hosam}@cs.ku.edu.kw

## ABSTRACT
Heterogeneous wireless sensor network consists of wireless sensor nodes with different abilities, such as different computing power and different initial energy. We present in this paper a new scheme for maximizing heterogeneous WSN lifetime. The proposed scheme employs two types of sensor nodes that are named (consistent with IEEE 802.15.4 standard) Full Function Device (FFD) and Reduced Function Device (RFD). The FFDs are the expensive sensor nodes with high power and computational capabilities compared to the RFDs which are cheap sensors with a limited power supply. The scheme divides the network into smaller sub-networks (regions) that are built from sectors and tracks. The objective of this research is to balance and reduce the communication load on RFDs, reduce the delay, and increase the connectivity and lifetime, by using a limited number of FFDs. We investigate the performance of our scheme via numerical simulation and compare it to other related schemes that are presented for homogeneous WSNs with chain topology, such as Pegasis, Epegasis and Chiron. In addition to extending the lifetime, our scheme also results in reducing the data transmission delay compared to the related schemes. Furthermore, the scheme increases the network security and reduces the RFDs power consumption by preventing the direct communication with the base station (BS). The FFD is the communication bridge between RFDs and BS. The FFD communicate with the BS using the one-hop approach or multi-hop approach through other FFDs.

## 1. INTRODUCTION
A Wireless Sensor Network (WSN) is a network that's composed of sensing, computing (if required) and communication elements called sensor nodes, that help in observing and reacting to events in a specified environment. Sensor nodes are used within self-organized networks that are mostly densely deployed inside the phenomenon or close to it. In WSN sensor nodes communicate (wirelessly) using radio frequency channel to acquire data from the physical world and transmit it (report it) to the BS (requesting destination) either through single or multi hop routing schemes.

WSNs are organized as homogenous or heterogeneous WSNs depending on the type of sensor nodes in the network. In general, homogenous WSNs are networks composed of the same type of sensor nodes, where in heterogeneous WSNs sensor nodes have different types. Choosing among these types of networks mostly depends on the application itself and the area type.

In IEEE 802.15.4 standard the names Full Function Device (FFD) and Reduced Function Device (RFD) are used to differentiate between the two type nodes in heterogeneous WSNs. FFDs are equipped with more power and can act as coordinators as well as common nodes, while RFDs have a limited power supply and very low processing capabilities. In the context of WSN, having different types of nodes may be useful for certain applications. For example, some applications need to build their network across areas that have different event occurrence (or sampling) rates, so higher event rates cause the sensors to consume more power than nodes at other areas which will cause them to die much earlier before the other nodes.

In this paper we propose a scheme for maximizing the lifetime of heterogeneous WSNs. We consider applications with a coverage area that can be approximated by a circular area. The area is divided into smaller areas (regions) built from sectors and tracks. At each region there is a chain that is composed of one FFD and a number of RFDS. The FFD is either at the end or middle of the chain. The RFDs are responsible to sense and report events in the region, where the FFDs are responsible to get the reports from the RFDs then transmit them to the BS either by one-hop or multi-hop approach.

The remainder of the paper is structured as follows. Section 2 covers the work related to our scheme. The network model, assumptions, sensor nodes overview and scheme structure are presented in section 3. Following that section 4 presents analysis and numerical simulation results of testing our model and comparisons with other models. Finally, Section 5 states our conclusions.

## 2. RELATED WORK

### 2.1 PEGASIS
Power Efficient Gathering in Sensor Information Systems (PEGASIS) is a chain based protocol [5]. In PEGASIS, nodes are arranged in a chain, where each node communicates only with its close neighbors. Transmitting the data to the BS is performed in rounds where at each round the data is collected from the nodes and transmitted directly to the BS by the Chain Head. The Chain head is a node that is elected randomly at each transmission round and it's responsible for receiving the reports from the nodes and transmitting them directly to the BS.

### 2.2 Enhanced PEGASIS
In 2007, a variation of PEGASIS was introduced by Jung et al. named as Enhanced PEGASIS (EPEGASIS) [4]. The main idea of this protocol is to consider the location of the BS to enhance its performance and to prolong the lifetime of the WSNs. In EPEGASIS the sensing area is centered at the BS into several centric cluster levels, where in each level there will be a chain of sensor nodes. EPEGASIS protocol consist of four processes; level assignment, chain construction in the level area, head node construction in chain, and data transmission. At level assignment process sensor nodes will be assigned their levels by the BS. The BS decides the number of levels depending on various parameters such as density on sensor network, number of nodes, or location of the BS. In chain construction process; for each cluster level a chain is constructed based on the greedy algorithm same as PEGASIS. In Head node construction process the EPEGASIS uses a mathematical way to elect the head node. Finally, at the data transmission process; the nodes at each level deliver the sensing data to the head node, same as PEGASIS strategy. After that, starting from the farthest level to the lowest level, a leader by leader data propagation task will be followed.

## 2.3 CHIRON

In 2009, energy efficient chain based hierarchical routing protocol, named as CHIRON was introduced [2]. The main idea of CHIRON protocol is to split the sensing area into smaller areas to have multiple shorter chains than EPEGASIS which will lead to reduce the data transmission delay and redundant path. Splitting the area is the responsibility of the BS by using Beam Star technique. In Beam Star technique the BS scan the area to broadcast control messages containing the ring and sector number. At each group nodes start constructing the chain by using greedy algorithm, same as PEGASIS. After the chain is constructed, the chain leader at each group is selected. The strategy that CHIRON follows in selecting the chain leader is that at initial round the chain leader is the farthest node from the BS then after initial round selecting the chain leader is based on the maximum value of residual energy of group nodes. The data transmission process is performed in two steps. At the beginning collecting the data in each group is performed same as PEGASIS, then starting from the furthest group the chain leaders collaboratively relay their aggregated sensing information to the BS.

## 3. THE PROPOSED PROTOCOL

In this section we propose a new scheme for maximizing the lifetime of heterogeneous WSNs. The scheme design is described as follows.

## 3.1 Network model and assumptions

Our proposed scheme was built on the following assumptions:

- The WSN consists of one stationary BS and several sensor nodes.
- The Area of deployment is a circle of radius R.
- WSN is heterogeneous; we have two types of nodes, FFD and RFD. FFDs have more computational and communicational power and initial energy (and maybe even connected to power lines).
- We have N RFDs and C FFDs, where N >> C.
- The FFDs are placed manually and the RFDs are uniformly distributed.
- All Sensor nodes have a very low-power Idle-listening mode.
- Sensor nodes are stationary and have unique IDs.
- The BS is placed in the center of the circle.
- The BS is not energy constrained and is equipped with a directional antenna with power control capability.
- A CDMA/CA MAC protocol is used by sensors.

## 3.2 Sensor nodes overview

This section presents an overview of the operation of sensor nodes used by the proposed scheme.

### 3.2.1 Node identification

In addition to the unique identification (ID) information (serial number) attached to each sensor node, each FFD and RFD stores the sector number and track number assigned to them by the BS.

### 3.2.2 Node Modes

Duty cycling technique is used to save the sensors battery. In duty cycling the sensor nodes switch from active mode(s) to a sleep mode(s) in order to save their battery power. We use the energy saving modes that was presented in [7] with some additional modes [1].

In the setup phase, the FFD and RFDs are in TR-On-Duty mode to build their topology, then after the topology is built the first round starts and the RFD switch to Listening-Duty mode. At the start of each round all the FFDs are in TR-On-Duty mode to receive the reports from the RFDs and send these reports to the BS. When an FFD sends the reports to the BS, then it switches to Off-Duty mode and wake when the next round starts using a Timer. The RFDs are always in Listening-Duty mode and wake only when they receive a packet. When a RFD receives a packet then it directly switches to TR-On-Duty mode to perform the required action then go back to the Listing-Duty mode. However, it can be switched to Sleep mode and waken up by a Timer at the beginning of the next round.

### 3.2.3 Node communication

FFDs can communicate only with the RFDs in their regions or with the other FFDs in the same sector or with the BS.

In the multi-hop approach the FFD need to know the predecessor and successor FFD in order to guarantee the correctness of the data path. Therefore, each FFD is equipped with a table which contains the predecessor and successor id's.

The predecessor is the previous FFD in the sector and the successor is the next FFD. Predecessor field for FFDs at last level is set to null. The successor for FFDs in the first level is the BS. In the one-hop the predecessor is set to null and the successor to the BS ID.

The RFDs have to know their neighbor's id's in order to communicate with them to deliver the data to the FFD. Each RFD is equipped with a table containing the ID of the predecessor and successor neighbor.

### 3.2.4 Message format

Each message have a header that contains ID, type, and routing information. The BS and nodes messages are described briefly as follows:

#### 3.2.4.1 BS

The base station has the responsibility to inform the RFDs about their location by scanning the whole network and broadcasting a control message containing the sector and the track number. This technique is similar to the technique used in the beam star approach [6] with some changes [1].

#### 3.2.4.2 Nodes

Table 1 shows the type of messages that are used by the scheme sensor nodes [1].

## 3.3 The proposed scheme

In this section we present the scheme structure in details. The scheme consists of four phases; setup phase, self-organization phase, data collection phase, and data transmission phase. The phases are described as follows:

### 3.3.1 Setup phase

The setup phase is composed of four steps: Deploy RFDs, RFDs position, Deploy FFDs, and Region head. In the first step, the RFDs are deployed to cover the geographic area. The deployment of RFDs follows a uniform distribution; therefore the expected number of RFDs in a unit area donated as ($\Gamma$) is given by (1):

$$\Gamma = \frac{N}{\pi R^2} \quad (1)$$

Where N is the number of RFDs and $\pi R^2$ is the circle area.

The scheme geographic circle area is divided into regions. The regions are formed by dividing the circle area into sectors ($n_s$) with equal angels ($\theta$) and tracks ($n_t$) (or levels) with equal widths(r). The sectors are ordered clockwise where the tracks start from the center to out.

The number of regions donated as M in the area is given by (2):

$$M = n_t * n_s \quad (2)$$

Furthermore, the number of FFDs is equal to M, since each region has only one FFD.

The number of tracks is the result of dividing the circle radius by the track width as shown in (3), where the number of sectors is the result of dividing $2\pi$ by theta as shown in (4):

$$n_t = \frac{R}{r} \quad , 0 < r \leq R \quad (3)$$

$$n_s = \frac{2\pi}{\theta} \quad , 0 < \theta \leq \pi \quad (4)$$

Since the network is divided into regions that are built from tracks and sectors, and the RFDs are uniformly distributed. Then the expected number of RFDs in a region at level i donated as ($\tau_i$) is given by (5):

$$\tau_i = \Gamma A_i \quad (5)$$

**Table 1. Sensor nodes messages**

| Message Type | Purpose | Type | Used by |
|---|---|---|---|
| Report | Data reported by RFDs | Unicast | FFD RFD |
| Req | Request for data (token packet) | Unicast | FFD RFD |
| D_Node | Message sends by a node to notify the BS that it will be dead soon. | Unicast | FFD RFD |
| Build_Chain | Build the chain in the topology | Unicast | FFD RFD |
| H_Region | Sent by the FFD to inform the RFDs that it's the region head | Broadcast (Region) | FFD |
| Chain_D | Sent by RFDs to inform the FFD that the build chain process is done | Unicast | RFD |
| Topology_D | Sent by FFDs to inform the BS that the build topology process is done | Unicast | FFD |

Where $A_i$ is the region area at track i. The region area size at level i ($A_i$) is:

$$A_i = \left(i - \frac{1}{2}\right) \theta r^2 \quad 1 \leq i \leq n_t \quad (6)$$

From (1), (5) and (6) the final form for the expected number of RFDs in a region at level *i* donated as $\tau_i$ is given by (7):

$$\tau_i = \frac{N \theta r^2}{\pi R^2}\left(i - \frac{1}{2}\right) \quad 1 \leq i \leq n_t \quad (7)$$

The second step of the setup phase is the RFDs position. The BS scans the area to inform the RFDs about their positions by sending control messages containing the sector and track number.

In step three; the FFDs are placed manually at the center of each region to be the region cluster head as shown in figure 1. The distance between any two consecutive FFDs at the same sector is r, where the distance between the first FFD and the BS is $\left(\frac{r}{2}\right)$. Therefore, the distance from an FFD at level i to the BS donated as $d_{BS}$ is equal to:

$$d_{BS} = \left(i - \frac{1}{2}\right)r \quad (8)$$

Finally, the fourth step of the setup phase: at each region the FFDs inform their RFDs that they are the head regions by sending messages that contain their ID's.

### 3.3.2 Self-organization phase

FFDs and RFDs collaborate to build the region topology (Chain), where the FFDs are responsible to start the building process. The FFD position would be either in the start of the chain or in the middle of the chain as shown in figure 2. When the FFD is placed in the start of the chain this means the FFD is connected to one chain (Chain1). Whereas it is connected to two chains when it's in the middle of the chain (Chain2).

The FFD starts building the chain by calculating the chain length using the Chains length algorithm in figure 3. After that the FFD will store the result in the Build_Chain message and send it to the closest RFD and wait for response. When an RFD receives the Build_Chain message then it will start building the chain with the other RFDs in the region using Build chain algorithm as shown in figure 4. After finishing constructing the chain the closest RFD will send a Chain_D message to the FFD, which will then send a Topology_D message to the BS to inform it that the topology was built. Finally, the BS start the first round after it has receives the Topology_D message from all FFDs.

### 3.3.3 Data collection phase

The collection process performed at each round, to collect the data and send it to the BS. As shown in figure 5, the FFD requests the data from the chain by sending a token packet and waits for the data.

### 3.3.4 Data transmission phase

The data transmission phase starts when the FFDs receive the data from their regions. The received data is sent to the BS either by one-hop or multi-hop approach. In the one-hop approach the FFD sends the received data to the BS directly, as shown in figure 6(a). Whereas in the multi-hop approach, the FFD sends its data with the predecessor FFD data in the same sector to the successor FFD towards the BS, as shown in figure 6(b).

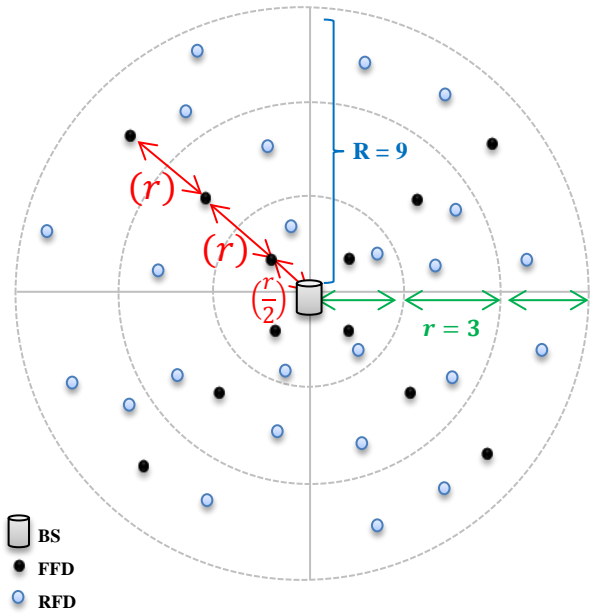

Figure 1. Scheme Structure

```
// This algorithm is used by RFD to build a chain in region at
level i (with a chain number h and length RDFs_Left).

1.  Min_Distance←R
2.  RFD[Current_RFD].Pre ← FFD_ID;
3.  RFD[Current_RFD].ch ← h
4.  DECREMENT  RFDs_Left
5.  WHILE RFDs_Left ! = 0
6.    For j ← 1 To τ_i
7.     IF  j != Current_RFD AND  RFD[j].Visit = 'N'  THEN
8.       Distance ← distance for Current_RFD to RFD[j]
9.       IF Distance < Min_Distance THEN
10.         Min_Distance ← Distance
11.         Next_RFD ← j, break;
12.      END IF
13.     END IF
14.     INCREMENT j
15.   END FOR
16.   RFD[Current_RFD].Suc ← Next_RFD
17.   RFD[Next_RFD].Visit ← 'Y'
18.   RFD[Next_RFD].Pre ← Current_RFD
19.   Current_RFD ← Next_RFD
20.   DECREMENT  RFDs_Left
21. END WHILE
22. RFD[Current_RFD].Suc ←NULL
```

Figure 4. Build chain Algorithm

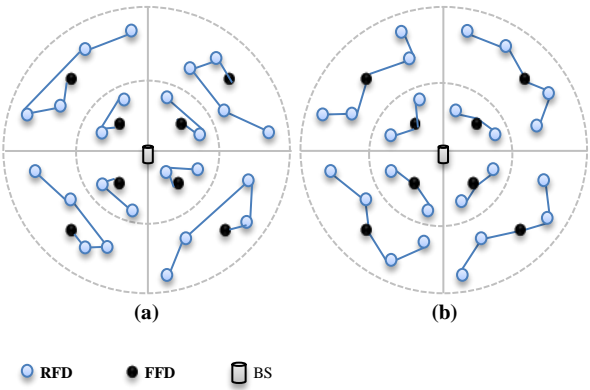

Figure 2. Chain Topology. (a) Chain1. (b) Chain2.

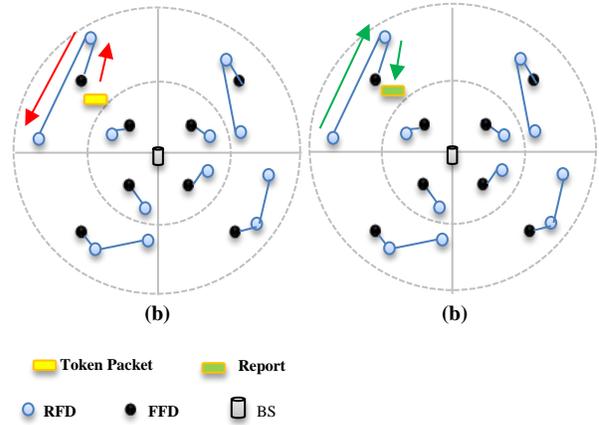

Figure 5. Data collection phase (Ex. Chain1). (a) FFD sends token packet to the RFDs asking for data. (b) RFDs send their reports to the FFD

```
// This algorithm is used by FFD to calculate the chain length in
the Chain topology with β chains.

1.  Chain_Length ← τ_i/β
2.  IF [Chain_Length] = Chain_Lenght  THEN
3.     /All chains have equal length
4.     RFDs_Left ←  Chain_Length
5.  ELSE
6.      Φ ← τ_i%β
7.      Φ chains with length:
8.      RFDs_Left ←  1 + [Chain_Length]
9.      β − Φ  chains with length:
10.     RFDs_Left ←  [Chain_Length]
11. END IF
```

Figure 3. Chains length Algorithm.

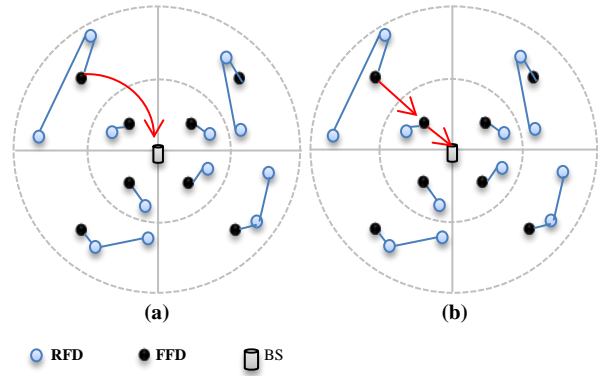

Figure 6. Data transmission Phase. (a) One-hop approach. (b) Multi-hop approach.

## 4. SIMULATION AND RESULTS

Our scheme performance has been tested numerically by creating equations that are built from energy models and by creating a simulator using C++ language [1].

### 4.1 Simulation environment and parameters

In our simulation we used one of the commonly used energy models used to formulate the energy dissipation of the FFDs and RFDs, the energy model is same as that used in [3]. The amount of consumed energy ($E_{T_x}$) to transmit k-bit packet with distance d meters is given by 9:

$$E_{T_x}(k,d) = E_{T_x-elec}(k) + E_{amp}(k,d)$$

$$E_{T_x}(k,d) = \begin{cases} k\,E_{elec} + k\,\epsilon_{fs}\,d^2 & \text{if } d < d_o \\ k\,E_{elec} + k\,\epsilon_{mp}\,d^4 & \text{if } d \geq d_o \end{cases} \quad (9)$$

In receiving k bits:

$$E_{R_x}(k) = E_{Rx-elec}(k) = kE_{elec} \quad (10)$$

The parameter values during simulation are assumed as follows:

$$E_{elec} = 50 \text{ nJ/bit}, \epsilon_{mp} = 0.0013 \text{ pJ/bit/m}^4,$$
$$\epsilon_{fs} = 10 \text{ pJ/bit/m}^2, d_o = \sqrt{\epsilon_{fs}/\epsilon_{mp}} \cong 87.7 \text{ m}$$

In this simulation we assume that the sensor nodes are deployed on a flat circle area with radius (R) 50 meters. The environment grid sensing area is 100X100m. The sensing radius for the FFDs and RFDs is 10 meters. The transmission range for sensor nodes is fixed. The transmissions range of the sink is 20 meters and could be increase to reach all the nodes in the network. The FFDs are equipped with a 100 Joules battery power, where the RFDs are equipped with a 10 Joules battery power. The sink has unlimited electric power. The minimum battery threshold for FFDs is 0.5 Joules and 0.05 Joules for RFDs.

Most model parameters came from the energy model, since the simulation model is mainly built from energy equations. In table 2 we summarized the main parameters of the simulation model that play an important role in the simulation process. The related models parameters values are presented in table 3.

### 4.2 Simulation results

In this section we study the effect of the number of sectors and tracks on the proposed scheme lifetime. Compare the proposed scheme models to the related models. The comparison is divided into three parts; network power consumption and lifetime, delay and cost.

**Table 2. Simulation Scheme parameters**

| Parameter | Symbol | Value |
|---|---|---|
| Theta | θ | 180° |
| Track width (level width) | r | 25 |
| Number of sectors | $n_s$ | 2 |
| Number of tracks | $n_t$ | 2 |
| Number of FFDs | C | 4 |
| Number of RFDs | N | 100 |
| Number of RFDs in level **i** | $N_i$ | – |
| Data message size | K | 2000 bits |

**Table 3: Related models parameters values**

| Model | Parameters values |
|---|---|
| Pegasis | N=100 (Number of sensor nodes) |
| | Area size= 100X100m |
| Epegasis | N=100 (Number of sensor nodes) |
| | Area: ( *Size* = 100X100m, *Radius* (R) =50 ) |
| | Number of tracks ($T_n$) =2 |
| | Track width (r) =25m |
| Chiron | N=100 |
| | Area: ( *Size*= 100X100m, *Radius* (R) =100, *Angle* ($\theta_{area}$)=90 ) |
| | Number of Levels ($L_n$)=2 |
| | Number of Sectors=2 |
| | Level width (r) =50m |
| | Sector angel($\theta_{sector}$)=45 |

*4.2.1 Effect of $n_s$ and $n_t$ on the proposed scheme*

The number of sectors and tracks decide the number of regions (i.e. number of FFDs) and RFDs in each region. Increasing $n_s$ or $n_t$ leads to increase the number of regions therefore decrease the number of RFDs in regions. Increasing the number of regions has two advantages. The first advantage is that the RFDs in a region communicate with less number of RFDs which means that the number of messages decreases therefore increase RFDs lifetime. The second advantage is that; increasing the $n_s$ or $n_t$ leads to increase the number of FFDs therefore decrease the load on the FFDs. Figure 7, 8, 9 and 10 shows the effect of $n_s$ and $n_t$ on the models lifetime in both approaches. Increasing the number of sectors and tracks leads to increase the lifetime of Chain1 and Chain2. The type of approach doesn't affect the models lifetime since the first dead node in the two models is RFD not FFD will be explained next.

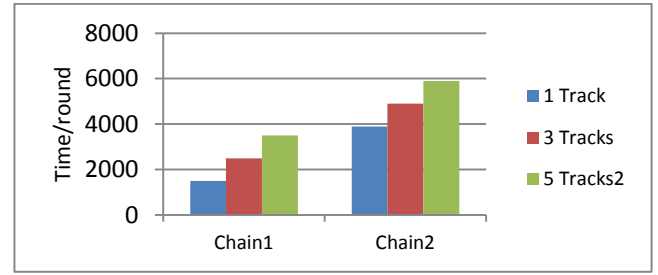

**Figure 7. The effect of $n_t$ on models lifetime (One-Hop approach)**

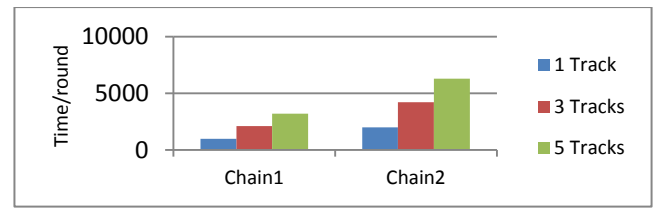

**Figure 8. The effect of $n_t$ on models lifetime (Multi-Hop approach)**

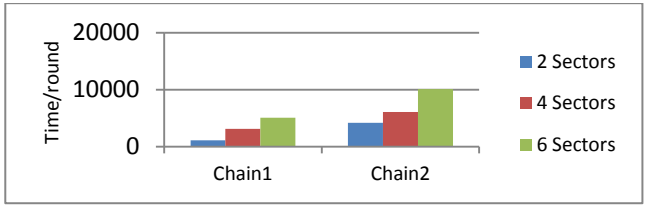

**Figure 9. The effect of $n_s$ on models lifetime (One-Hop approach)**

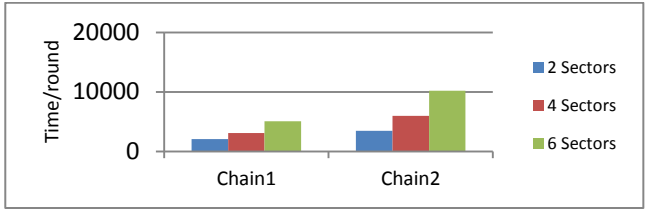

**Figure 10. The effect of $n_s$ on models lifetime (Multi-Hop approach)**

### 4.2.2 Power consumption and lifetime

As shown in figure 11, simulation results show that the Chain2 model is the model with highest lifetime comparing to the other models. This is related to the using of FFDs, the network structure and the FFDs position. FFD position decrease the data path and load on RFDs, which leads to increase the model lifetime.

Furthermore, the figure shows that the lifetime of CHIRON is greater than the Chain1 model. The CHIRON strategy (i.e. random head) leads to decrease the load on nodes by change the path at each round, which has the great advantage in prolonging the network lifetime.

Pegasis and Epegasis have low lifetime and this is related to the long chains comparing to the other models.

Simulation results shows that the first node dead in the Chain topology is the closest RFD to the FFD in the last level. Since this node receive and transmit the largest amount of data. Furthermore, simulation results show that the Chain1 and Chain1 model have the same lifetime in both approaches. This is an acceptable result since the only difference between both approaches is the FFD power consumption which is not affected in these models since their lifetime end when an RFD with the highest power consumption dies. The lifetime of Chain2 is greater than the lifetime of Chain1 because the load on the RFDs is different.

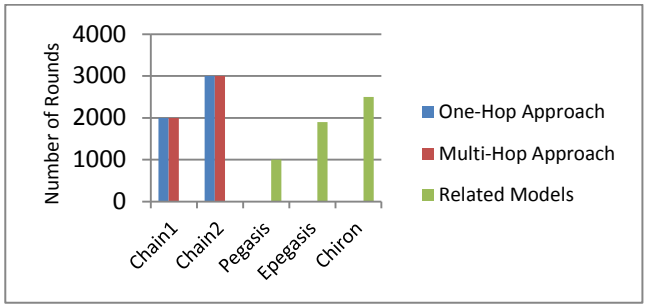

**Figure 11. Models lifetime**

### 4.2.3 Delay

The models delay is calculated according to the number of hops (unit of hops), where more hops means more delay. At each round the BS wait for all the packets which means it waits until the furthest packet arrived. The furthest packet is the packet with the longest path. The model with highest max-path is the model with highest delay. In the Chain1 and Chain2 models the max-path is the number of hops from the last RFD in a chain in region at last level to the FFD plus the number of hops from the FFD to the BS as shown in table 4. The number of hops from the RFD to the FFD is related to the number of RFDs in the chain. Therefore the total number of hops is equal to:

$$H = \begin{cases} \tau_{n_t} & \text{Chain1} \\ \left\lceil \frac{\tau_{n_t}}{2} \right\rceil & \text{Chain2} \end{cases}$$

Whereas the number of hops from the FFD to the BS is equal to one in the one-hop approach and equal to $n_t$ in the multi-hop approach. Therefore the maximum path is equal to:

$$\text{Max\_Path} = \begin{cases} H + 1 & \text{One Hop approach} \\ H + n_t & \text{Multi Hop approach} \end{cases}$$

The maximum path for the related models as shown in table 5 is calculated as follows:

In Pegasis the maximum packet path is when the head is placed at one of the chain ends. The packet with longest path goes through N-1 nodes until it reaches the head, which will send it to the BS by one hop. Therefore the total number of hops is N.

Epegasis the maximum packet path is when the head is placed at one of the chain ends in the last track. The maximum packet path is equal to the number of hops of a packet starting from the other chain end until it reaches the head ($\gamma_{T_n} - 1$) plus the number of tracks. The expected number of nodes in the last track is:

$$\gamma_{T_n} = \frac{N}{\text{circle area}} * \text{last region area}$$

$$= \frac{N}{\pi R^2} * \left( \pi r_i^2 - \pi r_{(i-1)}^2 \right)$$

$$= \frac{N\, r^2 (2i-1)}{R^2} \qquad \text{Where } i = T_n$$

The max-path for Epegasis is ($\gamma_{T_n} - 1$) plus the number of tracks, as shown in table 4.

**Table 4: The max-path for the three models**

| Model | Max-Path (One-Hop) | Max-Path (Multi-Hop) |
|---|---|---|
| Chain1 | $\tau_{n_t} + 1$ | $\tau_{n_t} + n_t$ |
| Chain2 | $\left\lceil \frac{\tau_{n_t}}{2} \right\rceil + 1$ | $\left\lceil \frac{\tau_{n_t}}{2} \right\rceil + n_t$ |

**Table 5. The max-path for the related models (path/hop)**

| Model | Max-Path |
|---|---|
| Pegasis | N |
| Epegasis | $(\gamma_{T_n} - 1) +$ Number of levels |
| Chiron | $(\omega_{L_n} - 1) +$ Number of regions |

In Chiron the maximum packet path is when the head is placed at one of the chain ends in region at last level, and the path from the head to the BS is to go through all heads. The expected number of nodes for a region at last level is:

$$\omega_{L_n} = \frac{N}{\text{sector area}} * \text{last region area}$$

$$= \frac{N}{(\frac{1}{2}\theta_{area}R^2)} * \left(i - \frac{1}{2}\right) \theta_{sector} r^2$$

$$= \frac{2N \theta_{sector} r^2}{\theta_{area} R^2} \left(i - \frac{1}{2}\right) \qquad \text{Where } i = L_n$$

As shown in figure 12, by using the max-path equations in table 4 and 5 the simulation results shows that the new models has less delay comparing to the related models. This result related to the models structure and the use of FFDs.

The network structure of Pegasis and Epegasis leads to have long chains comparing to the other models. Furthermore, the main reason that makes the Chain model faster than Chiron is the network structure since the chains length in Chain1 and Chain2 is shorter than in Chiron because we have more regions in Chain1 and Chain2 than in Chiron.

## 5. CONCLUTION

We proposed a new scheme for maximizing the lifetime of heterogeneous WSNs. The objective of this research is to discover a scheme that will reduce the power consumption of the RFDs resulting in prolonging the lifetime while minimizing the delay comparing to the related schemes (i.e. Pegasis, Epegasis and Chiron) through using FFDs. Other topologies such as star and star-chain are presented in [1]. Furthermore, some optimization techniques are used to optimize the number of FFDs in the area.

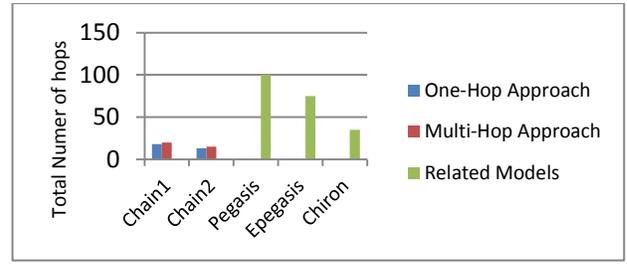

**Figure 12. Delay comparison between models**

Simulation results show that the proposed models are better than the related models in lifetime except the Chain1 model where the lifetime of Chiron model is better than Chain1. Furthermore, the proposed models have less delay comparing to the related models. In general the proposed models are better in lifetime and delay (but naturally not in cost) than the related models because of the model structure and the use of FFDs in their structure.

## 6. ACKNOWLEDGMENT

## 7. REFERENCES

[1] Reem Aldaihani. "A new secheme for maximizing the lietime of hetrogenous WSNs", M.S. thesis, Dept. Computer Sci., Kuwait Univ., 2011.

[2] Chen, K., Huang, J., & Hsiao, C. (2009). "Chiron: an energy-efficient chain-based hierarchical routing protocol in wireless sensor networks". Proceedings of the Wireless telecommunications symposium, wts'09 (pp. 183-187). USA: IEEE.

[3] Heinzelman, W., Chandrakasan, A., & Balakrishnan, H. (2002). An application-specific protocol architecture for wireless microsensor networks. IEEE Transactions On Wireless Communications, 1(4), 660-670.

[4] Jung, S.-M., Han, Y., & Chung, T.-M. (2007). "The concentric clustering scheme for efficient energy consumption in the pegasis". Proceedings of the The 9th international conference on advanced communication technology (pp. 260-265). Gangwon-Do: IEEE. 10.1109/ICACT.2007.358351.

[5] Lindsey, S., & Raghavendra, C. (2002). "Pegasis: power-efficient gathering in sensor information systems". Proceedings of the Ieee aerospace conference proceedings, 2002. (pp. 3-1125 - 3-1130 vol.3). IEEE. 10.1109/AERO.2002.1035242.

[6] Mao, S., & Hou, T. (2007). Beamstar: an edge-based approach to routing in wireless sensor networks. Transactions on Mobile Computing, 6(11), 1284-1296.

[7] Wang, L., & Xiao, Y. (2006). A survey of energy-efficient scheduling mechanisms in sensor networks. Mobile Networks and Applications, 11(5), 723-740.